\documentclass[twocolumn,showpacs,preprintnumbers,amsmath,amssymb]{revtex4}

\usepackage{graphicx}
\usepackage{dcolumn}
\usepackage{bm}

\begin{document}


\title{Influence of aging on fishtail-effect in pure and Ag-doped MG-YBCO}
\author{D. A. Lotnyk}
\email{dmitry.a.lotnik@univer.kharkov.ua}
\author{R. V. Vovk}
\author{M. A. Obolenskii}
\author{A. A. Zavgorodniy}
\affiliation{Physical department, V.N. Karazin Kharkov National University, 4 Svoboda Square, 61077 Kharkov, Ukraine.}
\author{J. Kov\'a\v{c}}
\author{V. Antal}
\author{M. Ka\v{n}uchov\'a}
\author{M. \v{S}ef\v{c}ikov\'a}
\author{P. Diko}
\affiliation{Materials Physics Laboratory, Institute of Experimental Physics, Slovak Academy of Sciences, Watsonova 47, 04001 Ko\v{s}ice, Slovakia}
\author{A. Feher}
\affiliation{Centre of Low Temperature Physics, P.J. \v{S}afarik University, Park Angelinum 9, 041 54 Ko\v{s}ice, Slovakia}
\date{\today}

\begin{abstract}
$M(B)$-curves were experimentally investigated. Fishtail-effect (FE) was observed in MG YBa$_2$Cu$_3$O$_{7-\delta}$ and YBa$_2$(Cu$_{1-x}$Ag$_x$)$_3$O$_{7-\delta}$ (at $x \approx$ 0.02) crystals in a wide temperature range 0.4~$<T/T_c<$~0.8 at the orientation of magnetic field $\textbf{H}\parallel c$. It was obtained that the influence of bulk pinning on FE is more effective at low temperatures while surface barriers dominates at high temperatures. The value $H_{max}$ for Ag-doped crystals is larger than for a pure one that due to the presence of additional pinning centers, above all on silver atoms.
\end{abstract}
\maketitle

\section{Introduction}
Since the discovery of high-temperature superconductors
(HTSCs), their engineering applications at liquid nitrogen
temperatures have drawn much attention.
The YBa$_2$Cu$_3$O$_{7-\delta}$ (YBCO) is one of the promising HTSCs materials
 for various technical applications. However, in ceramic materials the critical current density
($J_c$) is, due to the weak link effects, very low, rendering them unsuitable for applications. In
order to enhance $J_c$, the so-called melt-textured growth
(MG) process \cite{Jin88} has been developed, which can significantly
enhance $J_c$. One interesting phenomenon in the
MG HTSCs is the fishtail effect (FE) in $J_c$ as well as in
isothermal magnetic hysteresis loops ($M(H)$ curves) \cite{Muralidhar02, Koblischka00, Muralidhar03}.
As for the FE origin, some researchers propose the vortex
ordered phase to disordered phase transition \cite{Henderson96, Paltiel00, PaltielPRL00} to explain this interesting phenomenon, while others attribute it to the bulk pinning \cite{Muralidhar02, Koblischka00, Muralidhar03}. In the vortex phase transition
explanation, the FE exists both on a single $J_c(T)$ curve and
on a single $J_c(H)$ curve. However, in the bulk pinning explanation,
the FE appears on $M(H)$ curves whereas on a
single magnetization vs. temperature $M(T)$ curve it is seldom
reported.
In the previous work, the critical state model is used to
study the isothermal $M(H)$ curves \cite{Johansen97}. However, for
HTSCs, due to their operation at higher temperatures
and due to their small activation energy $U$, the flux creep is significant.
Hence, non-linear flux creep models were
developed. Meanwhile, the surface barrier \cite{Bean64}, which prevents
vortex entering ($H_{en}$) and exiting ($H_{ex}$) superconductors,
has strong effects on the irreversibility of
HTSCs. In \cite{Burlachkov93} pointed out that in the
case of a competition between bulk and surface pinning,
the initial relaxation is determined by the smaller pinning
energy between $U_{bulk}$ and $U_{surface}$ . Thus, for example,
if $U_{bulk}<U_{surface}$ then the bulk relaxation should
be observed first. In \cite{Zhang06} it was explained that the PE appears due to both bulk pinning and surface barriers therewith bulk pinning is more important at low temperatures while surface barriers at high temperatures. The goal of this work was to contribute to the
understanding of the FE origin in terms of significant influence of both bulk pinning and surface barriers.
\section{Experiment details and samples}
Undoped and Ag - doped YBCO bulk single-grain samples were fabricated by a top seeded melt-growth process (TSMG). Oxide powders YBa$_2$Cu$_3$O$_7$ + 0.25Y$_2$O$_3$ + 0.5 wt.$\symbol{37}$ CeO$_2$ + addition of Ag$_2$O in concentration determined by $x$ = 0.02 in YBa$_2$(Cu$_{1-x}$Ag$_x$)$_3$O$_{7-\delta}$ were milled in a friction mill and pressed into cylindrical pellets of 20~mm diameter. A
SmBa$_2$Cu$_3$O$_7$ type seed with geometrical dimensions 2$\times$2$\times$1~mm$^3$ was placed in the middle of the top surface
of the pellet. The samples were heated in the chamber furnace
to a sintering temperature, $T_{sint}$ = 950$^{\circ}$C, sintered for 24~h and than
heated to melting temperature, $T_m$ =1040$^{\circ}$C with dwell 9~h, cooled
to 1000$^{\circ}$C and than YBCO crystal was grown during slow cooling to 950$^{\circ}$C. As grown samples were cut in two halves along the \{100\} plane. The cut surface was grinded on SiC papers and fine polished
with alumina powder. The microstructure of the sample was analyzed
by an optical microscope in normal and polarized light \cite{Diko08}. The pureness of Ag$_2$O is 0.005 Wt$\symbol{37}$. \{110\} type of twins with twin spacing about 200~nm. Twin boundaries are parallel to the $c$-axis and formed almost 45$^{\circ}$ angle with $a$-axis. Small samples for oxygenation and magnetization measurements were cut from the $a$ - growth sector of the bulks \cite{Diko00a} at a distance of 1~mm from the seed (top surface of the pellets). The samples dimensions are shown in Table~\ref{tab:1}.

Two samples, pure crystal (S1) and Ag-doped crystal (S2), were used in the present work. Physical properties of the samples are presented in  Table~\ref{tab:1}.
\begin{table}[h]
\begin{tabular}{|c|c|c|c|c|} \hline
Sample & $T_c$, K & $\Delta T_c$, K & a(b)$\times$b(a)$\times$c, mm$^3$ & m, mg\\ \hline
S1 & 90.2-89.8 & 1.0 & 2$\times$1.8$\times$0.7 & 12.2\\ \hline
S2 & 91.3-91.1 & 1.5 & 1.7$\times$1.6$\times$0.8 & 17.98\\ \hline
\end{tabular}
\caption{\label{tab:1}Physical characteristics for samples YBa$_2$Cu$_3$O$_{7-\delta}$ (S1) and YBa$_2$(Cu$_{1-x}$Ag$_x$)$_3$O$_{7-\delta}$ (S2)}
\end{table}
\begin{figure}
\includegraphics[clip=true,width=3.2in]{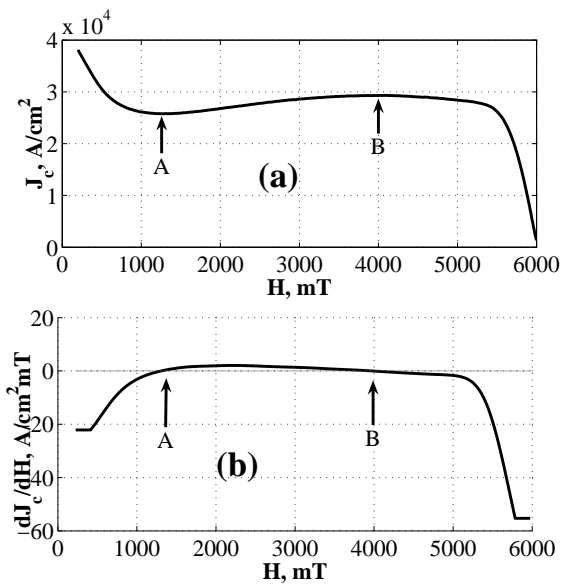}
\caption{\label{fig:1} Field dependences of critical current density $J_c$ (a) and derivative $dJ_c/dH$ (b) at temperature $T=$70~K for sample S1 right after annealing. Points A and B denotes positions of minimum and maximum respectively.}
\end{figure}
The measurements of isothermal hysteresis loops ($M(H)$ curves) at
various temperatures (20~K$<T<$85~K) were carried out in fields (-6~T$<H<$6~T) parallel to the c-axis by means of a commercial vibrating sample magnetometer (VSM). The field sweep rate was $dH/dt\approx 0.1$~T/min. The relaxation data were performed at each temperature at zero magnetic field (remanent magnetization) during 300~sec with $M>0$ and $M<0$. To obtain an additional parameter $\Delta T_c$ the resistivity $R(T)$ curves were measured on two samples (the first one is pure and the second one is Ag-doped) which were obtained from the same single-domain bulks as S1 and S2. Transport measurements were performed using a physical property measurement system (PPMS). The experiment was divided into two steps.  $M(H)$ curves were measured: a) right after annealing of samples in oxygen (during 240 hours at 400$^{\circ}$C); b) after long-term aging at the room temperature (six months), in which case almost all relaxation processes were finished. The first value for $T_c$ in Table~\ref{tab:1} corresponds to a) and the second value corresponds to b). The magnetic relaxation was measured for each curve at magnetic field $H=0$ during 300~sec.
\section{Model to analyze experimental data}
From the measured $M(H)$ curves the critical current density was calculated using the critical state model \cite{Bean62} that have been developed in \cite{Wiesinger92, Jirsa97}:
\begin{eqnarray}
{J_c = 2\Delta M\frac{\rho}{a^2\left(b-a/3\right)c}},
\label{eq:1}
\end{eqnarray}

where $\Delta M=M_1(H)-M_2(H)$ ($M_1(H)>$0, $M_2(H)<$0 are the values of magnetic moment at field $H$, $\rho$ is the weight of sample, a,b and c are geometric dimensions of sample ($b\geq a$). Dependences $J_c(H)$ (Fig.~\ref{fig:1}a for $T$=70~K) were calculated for the field range 0~T$<H<$6~T. To obtain the exact position of maximum on $J_c(H)$ curves the derivative $dJ_c/dH$ was calculated for each temperature (see example at Fig.~\ref{fig:1}b). Fig.~\ref{fig:1} shows positions of both minimum (point A) and maximum (point B) on J vs H curve, where derivative $dJ_c/dH=$0.
\begin{figure}
\includegraphics[clip=true,width=3.2in]{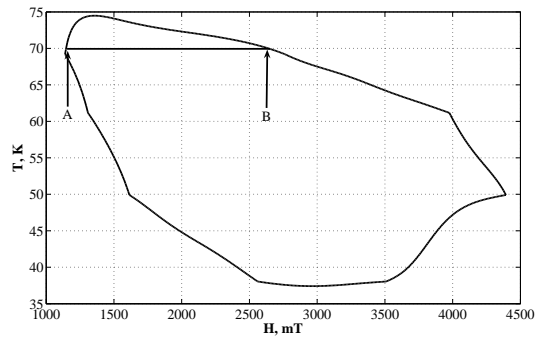}
\caption{\label{fig:2} Contour for $dJ_c/dH=0$ for sample S1 right after annealing. Points A and B coincide with points A and B in the Fig.~\ref{fig:1}}
\end{figure}
Then the 3D surface $T-H-dJ_c/dH$ was obtained. Intermediate curves were defined by the 3rd degree polynomial (spline-function). The contour for $dJ_c/dH=0$ is shown in Fig.~\ref{fig:2}. Points A and B correspond to those in Fig.~\ref{fig:1}. This method is useful to obtain position of maximum $J_{c,max}$ (or ensemble of points B) not only at given temperatures but also for intermediate temperatures.
\section{Results and discussions}
\begin{figure}
\includegraphics[clip=true,width=3.2in]{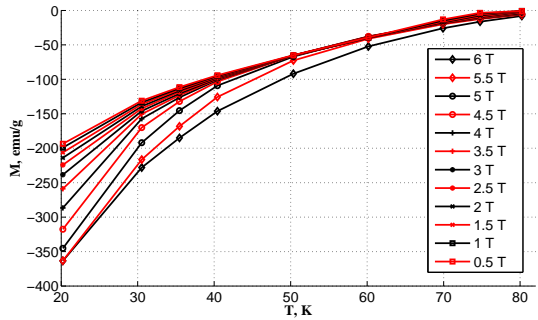}
\caption{\label{fig:3} M-T curves with ZFC process in fields from 0.5~T to 6~T for sample S1 6 month after annealing}
\end{figure}
\begin{figure}
\includegraphics[clip=true,width=3.2in]{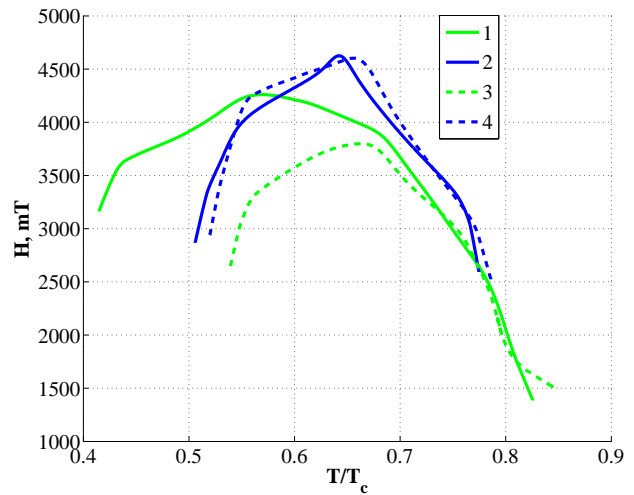}
\caption{\label{fig:5} Temperature-field dependences of $J_{c,max}$-position for samples S1 (green lines) and S2 (blue lines), right after annealing (curves 1,2) and after long-term aging (curves 3,4)}
\end{figure}

Fig.~\ref{fig:3} shows the zero-field-cooled (ZFC) $M-T$-curves at certain values of magnetic field (from 0.5~T to 6~T). As one can see magnetization monotonically changes with temperature and no apparent peak is visible in each ZFC $M-T$ curve. This is characteristic that the FE in $M-H$ curves cannot be simultaneously observed on a single ZFC $M-T$ curve, contrasting with the FE caused by order-disorder vortex matter transition. Hence, this can be used as a criterion to distinguish the FE originated from bulk pinning from the FE induced by order-disorder vortex matter transition \cite{Henderson96, Paltiel00, PaltielPRL00}. Measured magnetic relaxation provides a linear dependence in $M-lnt$ coordinates. The results for magnetic relaxation are shown in fig.~\ref{fig:5}.
\begin{figure}
\includegraphics[clip=true,width=3.2in]{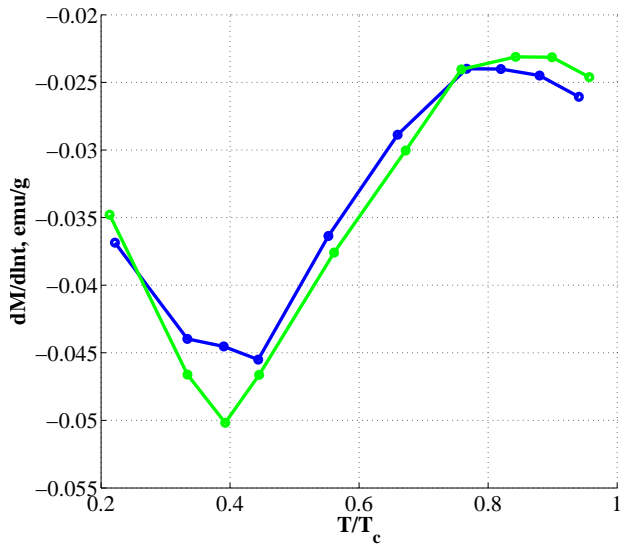}
\caption{\label{fig:4} Temperature dependencies of relaxation rate R=dM/dlnt for samples S1 (green) and S2 (blue) 6 months after annealing}
\end{figure}
The total magnetic moment $M=M_{surface}+M_{bulk}$. If surface activation energy $U_{surface}$ larger then the bulk one $U_{surface}\gg U_{bulk}$ the rate $R_{bulk}=dM_{bulk}/dlnt\gg dM_{surface}/dlnt$ \cite{Burlachkov93}, i.e. $dM/dlnt\approx dM_{bulk}/dlnt$. Otherwise, if $U_{bulk}\gg U_{surface}$ or $R_{surface}\gg R_{bulk}$ then $dM/dlnt\approx dM_{surface}/dlnt$. The crossover between these two regimes manifests in changes of slope $M-lnt$.

As one can see (fig.~\ref{fig:4}) relaxation rate $R=dM/dlnt$\cite{Burlachkov93} non-monotonically changes with temperature.
Temperature range can be divided into 3 parts:
\begin{enumerate}
\item At low temperatures (0.2$<T/T_c\lesssim$0.4) bulk pinning8 dominates over surface barriers, i.e. activation energy $U_{bulk}>U{surface}$ and total relaxation tare $R_{total}$ decreases. Position of minimum of relaxation rate $R$ for Ag-doped crystal shifts to a higher temperatures ($T/T_c=$0.4 for doping-free sample, $T/T_c=$0.45 for Ag-doped sample) as compared to doping-free sample. The bulk pinning for Ag-doped sample is stronger than for a doping-free one due to an additional pinning centers - Ag atoms \cite{Nakashima08}. That is why relaxation in Ag-doped sample due to $R_{surface}$ dominate to a higher temperatures then in undoped sample.
\item At intermediate temperatures (0.4$\lesssim T/T_c\lesssim$0.8) bulk pinning strongly decreased compared to surface barriers. It leads to a sufficient decreasing of activation energy $U_{bulk}$ and, hence, relaxation rate $R_{bulk}$ dominates over $R_{surface}$. As a result $R_{total}$ increases. It is significant to note that bulk pinning reduces with increasing temperature and at critical temperature disappears whereas surface barriers still finite. At the point $T=T_c$ the relaxation rate $R_{bulk}(T_c)=0$ .
\item At high temperatures ($T/T_c>$0.8) $R_{surface}$ dominates over $R_{bulk}$ because of $R_{bulk}\rightarrow$0 in the vicinity of $T_c$.
\end{enumerate}
From the model in previous chapter the $H-T$ dependences of $J_{c,max}$ were obtained (see Fig.~\ref{fig:5}).
As one can see from Table~\ref{tab:1} the value of the critical temperature $T_c$ for sample S1 changed from 90.2~K to 89.8~K and for sample S2 changed from 91.3~K to 91.1~K. This indicates the optimal oxygen annealing process and aging effects are not caused by changing in $\delta$ (especially for sample S2). Fig.~\ref{fig:5} shows that the position of $J_{c,max}$ significantly evolves after long-term aging. The maximum field $H_{max}$ for sample S1 decreases from 4.25~T to 3.8~T and the minimum temperature $T_{min}$ increases from 0.42 to 0.54 (curves 1 and 3 in Fig.~\ref{fig:5}). Such changes can be induced by the redistribution of pinning centers in the crystal or, in another words, bulk pinning changes during aging. Constancy of the maximum temperature $T_{max}$ for S1 corresponds to the influence of surface barriers at high temperatures. The aging effect for sample S2 is not so significant as for S1. It is related to the presence of considerable mechanical tensions in the sample (large value of an additional parameter $\Delta T_c=$1.5~K) that prevent the redistribution of pinning centers. Larger value of $H_{max}$ (for S2 as compared with S1) corresponds to the presence of additional pinning centers on silver atoms.

A possible reason of the decrease of $H_{max}$ and increase of $T_{min}$ for S1 is the decreasing number of dislocations. In another words, the crystal S1 becomes more homogeneous after aging \cite{Kupfer98}.

\section{Conclusions}
In summary, MG YBCO samples were investigated by measuring $M(B)$-curves. On these curves FE was observed in a wide temperature range. The origin of the FE provides in competition between surface barrier and bulk pinning at intermediate temperatures 0.4$\lesssim T/T_c\lesssim$0.8. It is confirmed in non-monotonically behavior of relaxation rate $R$, notably, increasing of $R$ at intermediate temperatures due to surface barriers. Bulk pinning influences parameters $H_{max}$ and $T_{min}$ while surface barrier influences $T_{max}$. Undoped YBCO is characterized by a significant aging effect which is manifested in a considerable change of $H_{max}$ and $T_{min}$. This corresponds to the change of bulk pinning. For Ag-doped YBCO there are no significant changes in parameters $H_{max}$, $T_{min}$ and $T_{max}$. Such difference is caused by the redistribution of pinning centers in undoped YBCO while Ag-doped YBCO is characterized by huge mechanical tensions which prevent redistribution.

\textit{Acknowledgments}

The work was partly supported by the Slovak Research and Development  Agency (No. APVV-0006-07) and  the Slovak Grant Agency VEGA (No.1/0159/09) The financial support of U.S.Steel - DZ Energetika Ko\v{s}ice is acknowledged. One of author (D.L.) is very thankful to the National Scholarship Program of Slovak Republic (SAIA) for the financial support during his stay in Slovakia.


\end{document}